\documentclass[twocolumn]{jpsj2} 
\title{Meta-Orbital Transition in Heavy-Fermion Systems:
Analysis by Dynamical Mean Field Theory and Self-Consistent
Renormalization Theory of Orbital Fluctuations}

\author{Kazumasa HATTORI}

\inst{Institute for Solid State Physics, University of Tokyo, 5-1-5 Kashiwanoha, Kashiwa, Chiba 277-8581, Japan
}

\abst{We investigate a two-orbital Anderson lattice model with Ising
 orbital intersite exchange interactions on the basis of a 
 dynamical mean field theory combined with the static mean field
 approximation of intersite orbital interactions. 
 Focusing on Ce-based heavy-fermion compounds, 
we examine the orbital crossover between two orbital states, when 
 the total f-electron number per site $n_{\rm f}$ is $\sim 1$.
We show that a ``meta-orbital'' transition, at which the occupancy of
 two orbitals changes steeply, occurs when the hybridization between
 the ground-state f-electron orbital and conduction electrons is 
smaller than that between the excited f-electron orbital and conduction
 electrons at low pressures. Near the meta-orbital critical end point, orbital 
fluctuations are enhanced and couple with charge fluctuations. 
A critical theory of meta-orbital fluctuations is also developed 
by applying the self-consistent renormalization theory of itinerant 
electron magnetism to orbital fluctuations.
The critical end point, first-order transition, and 
 crossover are described within Gaussian approximations of orbital
 fluctuations.  
We discuss the relevance of our results to CeAl$_2$, CeCu$_2$Si$_2$, 
 CeCu$_2$Ge$_2$, and related compounds, which all have low-lying
 crystalline-electric-field excited states. }

\kword{dynamical mean field theory, two-orbital Anderson lattice model,
self-consistent renormalization theory, critical end point, orbital fluctuations}

\begin{document}
\maketitle
\section{Introduction}
After the discovery of the unconventional superconductivity in 
CeCu$_2$Si$_2$\cite{Steglich}, various heavy-fermion compounds have
attracted great attention. They exhibit many 
interesting phenomena,\cite{Summary} such as quantum critical behaviors 
associated with the quantum critical point (QCP) of magnetic
phase transitions under pressure, magnetic field and chemical substitutions, 
non-Fermi liquid properties near the QCP, and various types of 
unconventional superconductivity. 
It is expected that the Cooper pairs in such heavy-fermion 
superconductors are mediated by magnetic fluctuations, since 
the superconductivity always occurs near the magnetic phase.

Over the past ten years, the possible nonmagnetic-fluctuation-mediated 
superconductivity at high pressures 
separated from the superconductivity near the magnetic QCP has attracted much attention in some heavy-fermion
compounds, such as CeCu$_2$Si$_2$, CeCu$_2$Ge$_2$, and related materials\cite{Jaccard1,Vargoz,Yuan,Holmes,Jaccard},
since magnetic fluctuations are not developed there.\cite{Fujiwara} Near the 
pressure where the superconducting transition temperature has a peak as a
function of pressure, electric resistivity shows a 
linear temperature $T$ dependence, with residual resistivity
being enhanced\cite{Jaccard, Yuan, Holmes}. As pressure increases, 
Kadowaki-Woods ratio gradually varies from the value for
strongly correlated systems to that for weakly correlated systems\cite{KWrelation}.

To explain superconductivity without magnetic fluctuations,
Onishi and Miyake\cite{Onishi} proposed that Ce-valence fluctuations
lead to 
 unconventional superconductivity at high pressures in these
systems on the basis of an extended periodic Anderson model (Ex-PAM). The existence
of the valence transition in the Ex-PAM was confirmed 
 by more elaborate numerical calculations
 later.\cite{Watanabe,Sugibayashi} 
The properties near the critical end point (CEP) of valence fluctuations
explain the resistivity\cite{Holmes,Maebashi} and changes in Kadowaki-Woods
 ratio.\cite{KWrelation}

Experimentally, when the f-electron charge per site varies, owing to the
electrostatic potential change around a Ce site, the lattice constant
should also vary correspondingly, and some of the phonon branches should
be affected by this change. However, no direct evidence 
 for the steep changes in Ce valence has been reported.

In this paper, we will focus on the crystalline-electric-field (CEF),
{\it i.e.}, orbital 
states in these systems. The materials mentioned above all have
low-energy excited CEF states. CEF excitations are observed
by inelastic neutron scattering experiments\cite{neutron1,neutron2}. 
Resistivity data also indicate the existence of excited
CEF states, showing a two-peak structure as a function of temperature $T$. 
The two peaks correspond to two different Kondo temperatures. 
More importantly, the two peaks merge as pressure
increases, and the pressure where the two peaks merge approximately 
coincides with
the pressure where superconducting transition temperature increases 
as mentioned above.\cite{Jaccard} The two-peak structure and its variation
as a function of pressure are well explained by the analysis of an
orbital-degenerate Anderson lattice model.\cite{Nishida}
Interestingly, these
behaviors are also observed in CeAl$_2$, which is a prototype of
heavy-fermion compounds, 
 at around 3 GPa, while superconductivity has not been
 observed.\cite{CeAl2-recent} A neutron 
scattering experiment revealed that there are also low-energy CEF
states in CeAl$_2$.\cite{CeAl2-neutron}

Thus, it is natural to consider that the orbital (CEF) fluctuations in these
systems play an important role in realizing the unconventional
superconductivity at high pressures. In this paper,
we will examine the orbital
variations as a function of pressure on the basis of a dynamical mean field
theory (DMFT)\cite{dmft} and show that a ``meta-orbital'' transition or
crossover occurs when hybridizations between conduction and f-electrons
depend on the orbital and are sufficiently small compared with the f-electron energy
level. 
We also find that the orbital fluctuations couple with
f-electron charge degrees of freedom, leading to the variation in  
f-electron occupancy when the orbital crossover or transition occurs. 
We will also try to construct an effective orbital fluctuation theory.

This paper is organized as follows. In \S \ref{sec-dmft}, a
two-orbital Anderson lattice model is analyzed on the basis of a dynamical mean field
theory with Wilson's numerical renormalization group method used as the impurity
solver. 
A theory of critical fluctuations near the critical end point of orbital 
fluctuations is developed in \S \ref{sec-scr} on the basis of the
self-consistent renormalization theory. 
Finally, \S \ref{sec-sum} shows the summary of the present paper.

\section{Two-orbital Anderson Lattice Model with Ising Anisotropic
 Orbital Interaction}\label{sec-dmft}
In this section, we investigate orbital fluctuations in 
a two-orbital Anderson lattice model with
Ising orbital intersite exchange interactions on the basis of the
DMFT\cite{dmft} combined with the static mean field
approximation of intersite exchange interactions. As a solver for
the effective impurity problem in the DMFT, we use Wilson's numerical
renormalization group (NRG) method, which is powerful for investigating 
zero-temperature properties of the impurity problem.  We will show the
zero-temperature properties of this system, including the variations in 
the orbital occupancy, the f-electron density of states, and the
zero-temperature phase diagram.  
\subsection{Model}
We investigate the periodic Anderson lattice model on a Bethe lattice with two 
localized f-orbitals ($\alpha=A$ or $B$) hybridizing with a single
conduction electron band.\cite{Nishida} We also include Ising orbital-orbital intersite
interactions in our model and the Hamiltonian is given as  
\begin{eqnarray}
H&=&\sum_{ij\sigma}(-t_{ij}-\mu\delta_{ij})c^{\dagger}_{i\sigma}c_{j\sigma}
 +\sum_{i\alpha\sigma}\Big[v_{\alpha}(c^{\dagger}_{i\sigma}f_{i\alpha\sigma}+{\rm
 h.c.})\nonumber\\
 &+&(\varepsilon_{f\alpha}-\mu)f^{\dagger}_{i\alpha\sigma}f_{i\alpha\sigma}\Big]
 +\sum_{i\alpha}U_{\alpha}n_{if\alpha\uparrow}n_{if\alpha\downarrow}\nonumber\\
 &+&U'\sum_{i\sigma\sigma'}n_{ifA\sigma}n_{ifB\sigma'}
 -\frac{J^z_o}{z_{\rm n.n.}} \sum_{\langle i,j\rangle}T_{iz}T_{jz}.\label{H}
\end{eqnarray}
Here, $c_{i\sigma}$ and $f_{i\alpha\sigma}$ represent the annihilation
operators for conduction and
f-electrons at site $i$, spin $\sigma$, and orbital $\alpha$,
respectively. The f-electron number and the orbital operator are defined as 
 $n_{if\alpha\sigma}\equiv f^{\dagger}_{i\alpha\sigma}f_{i\alpha\sigma}$
 and $T_{iz}\equiv \sum_{\sigma}(n_{ifA\sigma}-n_{ifB\sigma})$, respectively. 
$J_o^z$ is the Ising orbital interaction between the nearest neighbor
sites, which is assumed as $J^z_o>0$, and $z_{\rm n.n.}$ is the number
of 
nearest-neighbor sites. Since, in the DMFT, 
intersite correlations are neglected, we introduce this
orbital interaction phenomenologically in order to take into account the
correlations. 
Orbital interactions are generated by a fourth-order
perturbation with respect to $v_{\alpha}$, starting from the localized 
limit in a two-orbital Anderson lattice model, as shown in Appendix \ref{Kondolimit}. 
 Hamiltonian (\ref{H}) includes only one of the many interactions generated
 in fourth-order perturbations, such as spin-spin exchange and
 interactions between spin-orbital degrees. Since we are interested in
 the orbital fluctuations in a paramagnetic phase, we restrict ourselves
 to take into account the Ising orbital interaction in this paper.
 The notations used for other parameters are conventional.  

Throughout this section, 
we set the nearest-neighbor hopping $t=1/2$, which
corresponds to half of the bandwidth $D=2t=1$, and 
we concentrate on analyzing zero-temperature $T=0$
properties of Hamiltonian (\ref{H}) and fix the total
electron number per site $n=1.8$, and $\varepsilon_{fA}=-0.5$ 
in the unit of $D$. As for the value of $J_o^z$, we will mainly discuss  
the case of $J^z_o=0.02$ while briefly showing the results for
$J_o^z=0$ as a reference. 
Since we are interested in Ce-based heavy-fermion compounds, 
we analyze the f$^1$ configuration,
{\it i.e.,} the configuration with
$n_{\rm f}\equiv\sum_{\alpha\sigma}n_{i\alpha\sigma}^f\simeq 1$. 
This is the reason why spin exchange interactions between $A$- and
$B$- orbitals are absent in Hamiltonian (\ref{H}), since they are
irrelevant in the f$^1$ configuration.  
In the following, $U_\alpha$ and $U'$ are set to be infinite 
in order to project out f$^2$-f$^4$ configurations. 

\subsection{Analysis by the dynamical mean field theory} 
In the DMFT, a lattice model is mapped to an effective impurity
 one with an effective conduction electron bath.\cite{dmft} In our case, the
effective impurity model is an impurity Anderson model with three localized
orbitals hybridizing with single-band conduction electrons. In this
paper, the intersite orbital interaction is approximated in the
mean field level with a uniform amplitude. The effective impurity model is given as 
\begin{eqnarray}
H_{\rm eff}\!\!\!\!\!\!&=&\!\!\!\!\!\!\sum_{k\sigma}\Big[\tilde{\epsilon}_k  a^{\dagger}_{k\sigma}a_{k\sigma}
+(\frac{\tilde{v}_k}{\sqrt{N}} a^{\dagger}_{k\sigma}c_{\sigma}+{\rm h.c.})\Big]
-\mu\sum_{\sigma}c^{\dagger}_{\sigma}c_{\sigma}\nonumber\\
\!\!\!\!\!\!&+&\!\!\!\!\!\!\sum_{\alpha\sigma}\Big[v_{\alpha}(c^{\dagger}_{\sigma}f_{\alpha\sigma}+{\rm
 h.c.})+(\bar{\varepsilon}_{f\alpha}-\mu)f^{\dagger}_{\alpha\sigma}f_{\alpha\sigma}\Big]\nonumber\\
\!\!\!\!\!\! &+&\!\!\!\!\!\!\sum_{\alpha}U_{\alpha}n_{f\alpha\uparrow}n_{f\alpha\downarrow}
 +U'\sum_{\sigma\sigma'}n_{fA\sigma}n_{fB\sigma'}.\label{Himp}
\end{eqnarray}
Here, $\bar{\varepsilon}_{f\alpha}$ includes the contributions of the
mean field value of $T_{iz}$:
$\bar{\varepsilon}_{fA}=\varepsilon_{fA}-J_o^z\langle T_{z}\rangle$ and 
$\bar{\varepsilon}_{fB}=\varepsilon_{fB}+J_o^z\langle T_{z}\rangle$, where
$\langle \cdots \rangle$ indicates the expectation value and $\langle
 T_z \rangle = \langle T_{iz} \rangle$. The operators
at the impurity site are defined as $f_{0\alpha\sigma}\equiv
f_{\alpha\sigma}$ and $c_{0\sigma}\equiv c_{\sigma}$. The effective bath 
is represented by the bath conduction electron $a^{\dagger}_{k\sigma}$
hybridizing only with $c_{\sigma}$. The dispersion and hybridization 
of the bath conduction electrons are characterized by
 $\tilde{\epsilon}_k$ and  $\tilde{v}_k$, respectively, which are calculated
 self-consistently in the DMFT.
Although one can integrate both $a_{k\sigma}$ and $c_{\sigma}$ in a 
path-integral formulation, 
it is useful to employ the form (\ref{Himp}) for NRG calculations.

To solve
 the impurity problem in the DMFT, we employ Wilson's NRG
 method\cite{Wilson,Bulla}, 
which is powerful for investigating 
the low-energy properties of the impurity problem. 
To calculate one-particle Green's functions correctly, 
we use the full-density-matrix NRG developed recently\cite{FD-NRG,Peters}. 
In the Bethe lattice, the self-consistent equation of the DMFT reads
\begin{eqnarray}
\Delta(\omega)=\frac{1}{N}\sum_{k}\frac{|\tilde{v}_k|^{2}}{\omega-\tilde{\epsilon}_k}=\frac{1}{4}G_c(\omega),
\end{eqnarray}
where $G_c(\omega)=-i\int_{0}^{\infty}dte^{-i\omega t} \langle T
\{c_{\sigma}(t),c^{\dagger}_{\sigma}(0)\}\rangle$ being the local retarded 
Green's function for conduction electrons.
In the NRG calculations in the DMFT self-consistent loops, 
we use the logarithmic descretization parameter $\Lambda=1.8$ and 
mainly keep 300 low-energy states 
at each NRG iteration.\cite{Wilson}  We also confirm that the
results do not change when we increase the number of retained states 
up to 700.

\begin{figure}[b!]
	\begin{center}
    \includegraphics[width=0.5\textwidth]{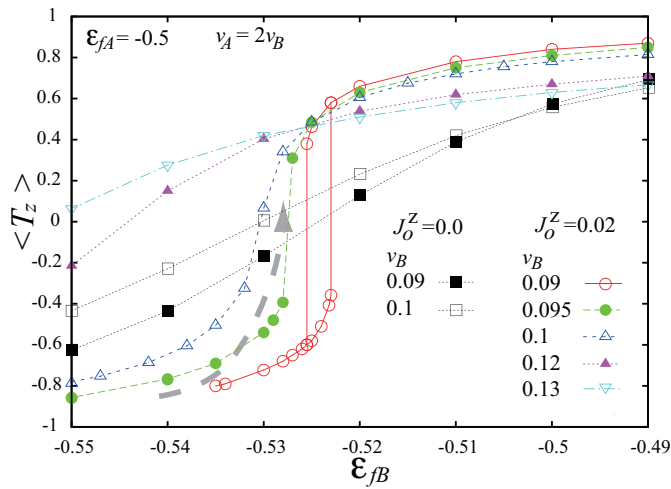}
\end{center}
\caption{(Color online) $\langle T_z\rangle $ vs $\varepsilon_{fB}$ for several values of
 $v_B=0.5v_A$, $\varepsilon_{fA}=-0.5$, and $J_o^z=0.0\ {\rm and}\ 0.02$, and the filling is
 fixed to be $n=1.8$. The arrow indicates the typical variation in 
 $\langle T_z\rangle$ for $J_o^z=0.02$ 
as a function of pressure for Ce-based heavy-fermion systems.}
\label{fig-Tz}
\end{figure}
\subsubsection{Meta-orbital transition}
In this subsection, we will show the numerical results of DMFT+NRG
calculations. We will demonstrate that a meta-orbital transition and
crossover 
occur in this system. We will also demonstrate that the orbital degrees of
freedom couple with the f-electron charge ones. 

Figure \ref{fig-Tz} shows the $\varepsilon_{f B}$ dependence of
orbital occupancy $\langle T_z\rangle$ for seven sets of $v_B$ with
 fixed $v_A=2v_B$. As $\varepsilon_{fB}$ decreases,  
$\langle T_z\rangle$ decreases from $\sim 1$ to $\sim -1$, indicating 
that the dominant orbital occupation changes from the $A$-orbital to 
the $B$-orbital. 
The variation in $\langle T_z\rangle$ as a function of
$\varepsilon_{fB}$ is
gradual when $v_B(=0.5v_A)$ is large, while it becomes steeper 
as $v_B$ decreases. Finite $J_o^z$ also enhances this steep change. 
Orbital occupancies for $J_o^z=0.0$ with $v_B=0.1$ (squares) and
$v_B=0.09$ (filled squares) 
change smoothly, while those for $J_o^z=0.02$ with the same parameters
show much steeper changes.

 A CEP is located between $v_B=0.095$ and 0.1 for $J_o^z=0.02$.  
For $v_B=0.09$, a
first-order transition occurs. One can see
a clear hysteresis, as shown in Fig. \ref{fig-Tz}.
 For $v_B=0.095$, although we have not detected a hysteresis in our
 calculation, it seems that there is a first-order transition.

These behaviors are
 analogues of meta-magnetism. In the present system, $\varepsilon_{fB}$
and $\langle T_z\rangle$ 
correspond to the magnetic field and magnetization in the language of 
meta-magnetism, respectively. This meta-orbital transition is expected to
occur in systems with several low-energy CEF 
states as the pressure increases. When the parameters are tuned, 
the CEP is realized.
In Fig. \ref{fig-Tz}, we also show
the expected variations in $v_B$, $\varepsilon_{fB}$, and $\langle T_z\rangle$ 
in Ce-based compounds, with increasing pressure, 
indicated by an arrow for $J^z_o=0.02$. Note that, 
as the pressure increases, both the hybridization and f-electron energy
levels are expected to increase in Ce-based heavy-fermion compounds. 

\begin{figure}[t!]
	\begin{center}
    \includegraphics[width=0.5\textwidth]{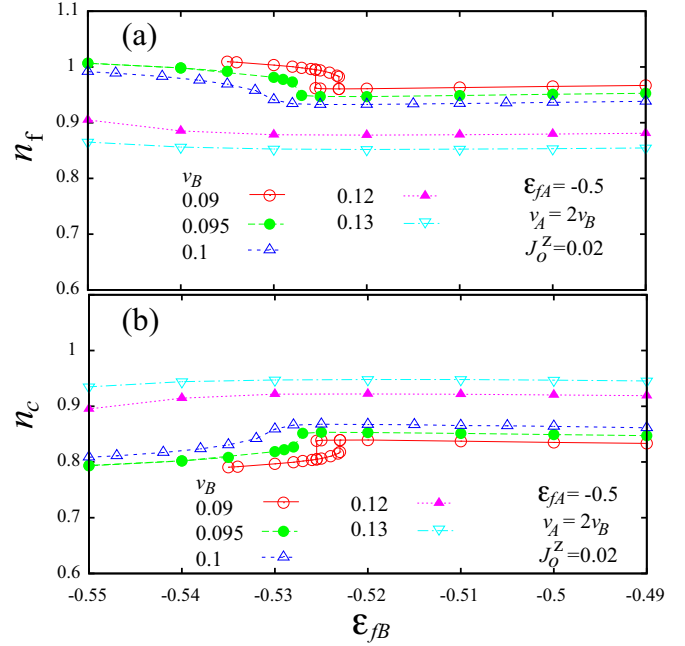}
\end{center}
\caption{(Color online) (a) $n_{\rm f}$ vs $\varepsilon_{fB}$ and (b) $n_c$ vs $\varepsilon_{fB}$ 
for several values of $v_B=0.5v_A$. The other parameters are the same as
 those in Fig. \ref{fig-Tz}.}
\label{fig-Nf-Nc}
\end{figure}

Figure \ref{fig-Nf-Nc} shows the $\varepsilon_{fB}$ dependences of the
total f-electron number $n_{\rm f}$ and the conduction electron number
$n_c$ with fixed $n=n_{\rm f}+n_c=1.8$.  
 Note that, for $v_B\le 0.1$, $n_{\rm f}$ is
close to 1 for a small $\varepsilon_{fB}$, while $n_{\rm f}\simeq 0.9$-$0.95$
for a large $\varepsilon_{fB}$. This is because $v_B<v_A$. When
$\varepsilon_{fB}$ is 
sufficiently small, 
 for example, $\varepsilon_{fB}=-0.55$ and $v_B=0.095$, $\langle T_z\rangle\sim
-0.9$, as shown in Fig. \ref{fig-Tz}, which indicates that 
the f-electron occupies essentially the $B$-orbital. On the
other hand, when $\varepsilon_{fB}$ is large, the f-electron occupies
mainly the $A$-orbital.
In these two limiting cases, it is 
sufficient to consider a one-orbital Anderson lattice model. We
obtain a low Kondo temperature $T_K$ that leads to $n_{\rm f}\sim 1$ for the small
$\varepsilon_{fB}$ with the relevant hybridization $v_B<v_A$, while 
a high $T_K$ that leads to $n_{\rm f}<1$ for the large $\varepsilon_{fB}$ with the relevant
hybridization $v_A>v_B$. 

As for the symmetry, the orbital, f-electron number, and
conduction electron number belong to the same irreducible
representation (scalar). Thus, as shown in Fig. \ref{fig-Nf-Nc},  
when the meta-orbital transition or crossover occurs, 
the f-electron number changes accordingly. 
We note that, across the transition and the crossover, the
f-electron number is still close to 1, when $v_A$ and $v_B$ are not very 
large as expected in heavy-fermion systems. This point contrasts the
results of the valence transition driven by the Coulomb repulsion between
the f- and conduction electrons, $U_{\rm fc}$, in the Ex-PAM, where a large variation in
f-electron number occurs, owing to the large $U_{\rm fc}$ assumed\cite{Onishi,Watanabe,Sugibayashi}.

\subsubsection{Variation in density of states}
\begin{figure}[t!]
	\begin{center}
    \includegraphics[width=0.5\textwidth]{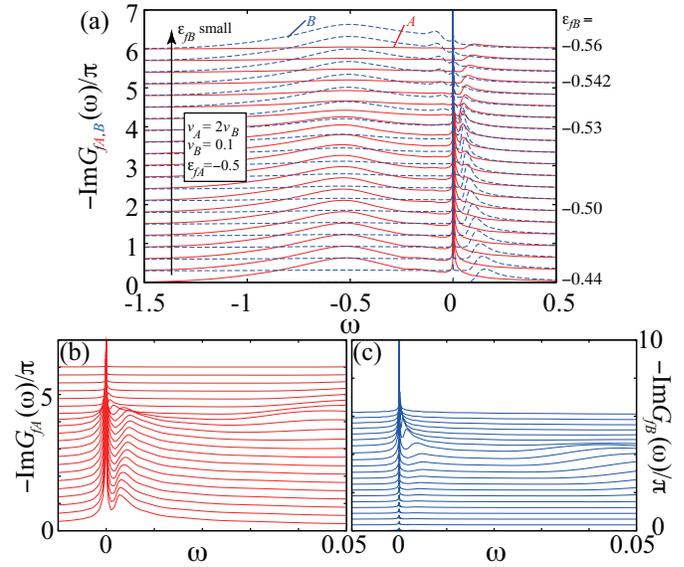}
\end{center}
\caption{(Color online) (a) f-electron density of states vs $\omega$ for $v_B=0.1$
 and  $v_A=0.2$ for several values of $\varepsilon_{fB}$. Full and dashed
 lines represent the DOS for the $A$-
 and $B$-orbitals, respectively. The values of $\varepsilon_{fB}$ decrease
 from bottom to top as $\varepsilon_{fB}=-0.44$, $-0.45$, $-0.46$, $-0.47$,
 $-0.48$, $-0.49$, $-0.50$, $-0.505$, $-0.51$, $-0.515$, $-0.52$, $-0.525$,
$-0.528$, $-0.53$, $-0.532$, $-0.535$, $-0.538$, $-0.542$, $-0.547$,
 $-0.555$, and $-0.56$. Each line is shifted by
 0.3. The other parameters are the same as those in Fig. \ref{fig-Tz}. (b)
 Low-energy structure of $A$-orbital density of states. (c) Low-energy
 structure of $B$-orbital density of states. }
\label{fig-ImGab}
\end{figure}

Figure \ref{fig-ImGab} shows the density of states (DOS) of f-electrons
$-{\rm Im}G_{fA,B}(\omega)/\pi$ for $v_B=0.1=0.5v_A$ and several values of
$\varepsilon_{fB}$. Here, $G_{fA,(B)}(\omega)$ represents the local 
$A$($B$)-orbital
f-electron retarded Green's function with the energy $\omega$.
In addition to the lower Hubbard-like peak located at the energy $\omega\sim\varepsilon_{fA,B}\sim -0.5$,
there are several peaks in the low-energy region, reflecting the CEF excited
states. The low-energy peak near $\omega\sim 0.2$ 
in the DOS of the $B$-orbital for $\varepsilon_{fB}=-0.44$ 
moves to a lower energy as $\varepsilon_{fB}$ decreases,
while a peak appears at a  positive energy in the DOS of the $A$-orbital as 
$\varepsilon_{fB}$ decreases,  
reflecting the meta-orbital
crossover discussed previously. 
There is a sharp peak near $\omega=0$ in $-$Im$G_{fA(B)}/\pi$. One can
also see that there is a hybridization gap like structure for the $A(B)$-orbital
DOS when $\varepsilon_{fB}$ is larger (smaller) than approximately
$\varepsilon_{fB}\sim -0.53$, as
shown in Figs. \ref{fig-ImGab}(b) and \ref{fig-ImGab}(c). 

\begin{figure}[t!]
	\begin{center}
    \includegraphics[width=0.5\textwidth]{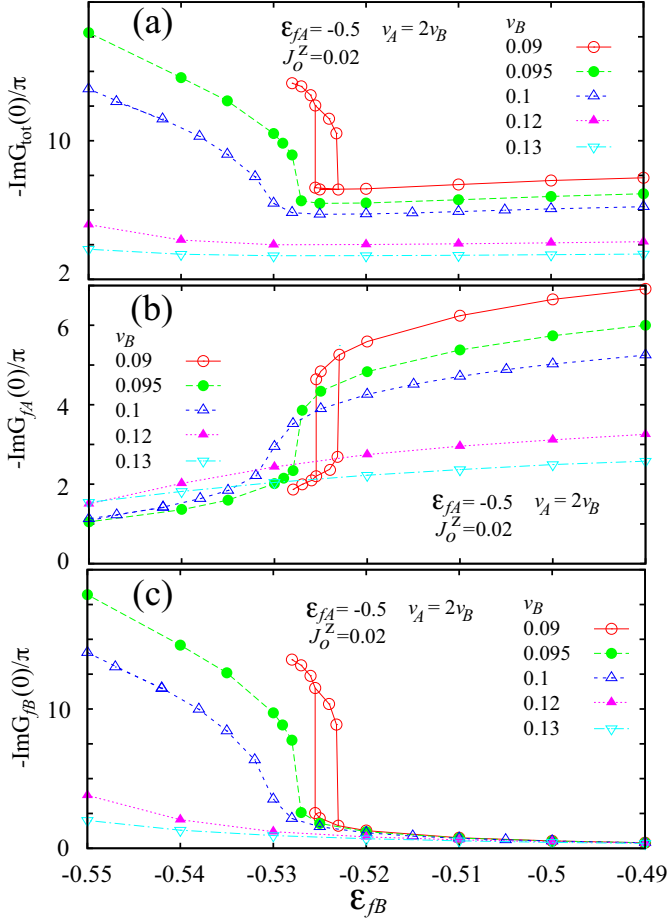}
\end{center}
\caption{(Color online) f-electron density of states at the Fermi level ($\omega=0$) vs $\varepsilon_{fB}$. (a) Total
 density of states $-$Im$G_{\rm tot}(0)/\pi$, (b)
 $-$Im$G_{fA}(0)/\pi$, and (c) $-$Im$G_{fB}(0)/\pi$. 
The other parameters are the same as those in Fig. \ref{fig-Tz}.}
\label{fig-dos}
\end{figure}

To investigate the low-energy DOS in more detail, 
we show the DOS at
$\omega=0$ as a function of $\varepsilon_{fB}$ 
in Fig. \ref{fig-dos}, where we define $G_{\rm
tot}=G_{fA}+G_{fB}+G_{c}$. 
Since Im$G_c$ does not show notable changes, we do not
show it. The DOS of the $A$-orbital at $\omega=0$ is large 
when $\varepsilon_{fB}$ is large, while it is small when 
$\varepsilon_{fB}$ is small. Correspondingly, the DOS of the 
$B$-orbital at $\omega=0$ 
is large for a small $\varepsilon_{fB}$ and small for a 
large $\varepsilon_{fB}$. In Fig. \ref{fig-dos}, one can also see a clear
hysteresis in the DOS at $\omega=0$ for $v_B=0.09$.

If we employ the variations in hybridization
and f-electron level as pressure increases as shown by the arrow in
Fig. \ref{fig-Tz}, 
we observe that the DOS at the Fermi level decreases as pressure
increases. The steepness of this variation depends on the distance to 
the CEP from the parameter path that corresponds to the physical
pressure.

\subsubsection{Phase diagram} \label{sec-phase}
\begin{figure}[t!]
	\begin{center}
    \includegraphics[width=0.5\textwidth]{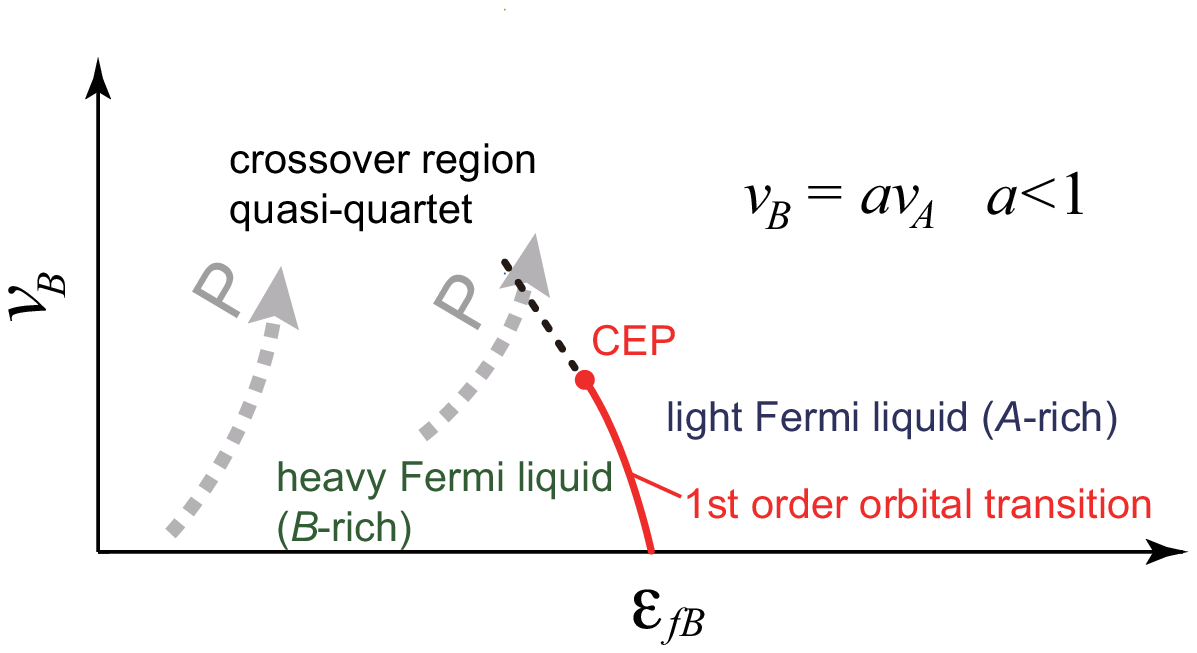}
\end{center}
\caption{(Color online) Schematic ground state phase diagram in $v_B$-$\varepsilon_{fB}$ plane
 for $v_B=av_A$ with fixed constant $a<1$. Arrows
 indicate examples of the parameter changes as the physical
 pressure $P$ increases. The dotted line indicates the orbital crossover line where
 $\langle T_z\rangle\sim 0$. The critical end point is represented by
 a filled circle and the first-order transition line is indicted by a solid
 line. The hysteresis related to the first-order transition is not
 indicated. The left side of the first-order transition corresponds to
 the state with a larger effective mass due to the smaller hybridization
 $v_B$, while the right side of the first-order transition corresponds to
 that with a smaller effective mass, denoted as light Fermi liquid. }
\label{fig-phaseDMFT}
\end{figure}

To summarize this section, we show the schematic ground state phase
diagram of the two-orbital Anderson model with two different
hybridizations ($v_B<v_A$) 
and ferro-orbital intersite interactions in Fig. \ref{fig-phaseDMFT}.
 As discussed above, we observe a first-order transition line and its 
CEP. In Fig. \ref{fig-phaseDMFT}, the ``$A(B)$-rich'' region indicates the
region with a larger f-electron occupancy in the $A(B)$-orbital. 
 The ``$A$-rich'' region is located in the larger $\varepsilon_{fB}$ region, where 
the effective mass is smaller than that in the ``$B$-rich'' region, 
owing to the large $v_A>v_B$. We also call it ``light'' Fermi liquid
instead of heavy Fermi liquid.
This light Fermi liquid can be regarded as the state with larger
 valence fluctuations than the $B$-rich state, since $v_A>v_B$. 
Near the CEP, f-electron charge 
fluctuations and also conduction electron ones are enhanced. 
The effective mass also 
changes steeply near the CEP, reflecting the variation
in orbital occupancy. 

Since our analysis is based on the DMFT and the mean field approximation for 
intersite orbital interactions $J_o^z$, we overestimate the effects of
$J_o^z$. 
 As shown in Fig. \ref{fig-Tz}, the 
$\varepsilon_{fB}$ dependence of $\langle T_z\rangle$ for $J_o^z=0$ is
much weaker than that for $J_o^z=0.02$. We have not carried out DMFT+NRG 
calculations for a very small $v_B<0.09$, since very small hybridizations
make the NRG calculation difficult.
When $J_o^z\to 0$, whether the first-order transition occurs or not is
beyond the scope of the present analysis. 
However, it is noted that when
one of the hybridizations is zero, there must be a first-order
transition even when $J_o^z=0$. 
To clarify whether this first-order
transition exists in the region where the hybridization is finite,
more elaborate numerical calculations that can take into account the
intersite correlations are needed.

As for magnetic instabilities, 
it is expected that magnetic phases will appear in
the phase diagram, especially in the region of the smaller $v_B$ and
$\varepsilon_{fB}$, corresponding to the low-pressure region in typical
Ce-based heavy-fermion systems. Determining the phase diagram
including the magnetic phases and their magnetic structures is beyond
the scope of the present paper and we leave it as a future study.
It is also interesting to investigate the case that the magnetic
transition and the orbital CEP occur simultaneously. Such a situation
might lead to a new universality class of magnetic phase transitions.

\section{Self-Consistent Renormalization Theory of Nonmagnetic
 Orbital Fluctuations} \label{sec-scr}
In this section, we will develop a self-consistent renormalization
theory for the 
orbital fluctuations discussed in \S \ref{sec-dmft} on the basis of the DMFT. 
 Since the DMFT does not take into account the long wavelength
fluctuations, we use the one-loop self-consistent approximation scheme 
of the fluctuations known as the self-consistent renormalization (SCR) theory
developed by Moriya and Kawabata for itinerant electron 
magnetism\cite{Moriya1,Moriya2,MoriyaBook}. When we apply the SCR theory to
the nonmagnetic orbital fluctuations, we need to take into account the temperature
dependence of the order parameter itself and the temperature dependence 
{\it originating from the order parameter in the mode-coupling corrections}. 
Our discussion in this section leads to results similar  to those of the 
SCR theory for meta-magnetic fluctuations\cite{Metamag} 
and Kondo-volume-collapse transition in heavy-fermion 
systems\cite{KVC}. 
It is noted that our formulation naturally describes both the order
parameter and the mode-coupling corrections in an equal footing based on a variational principle. 
We will discuss the orbital fluctuations in three-dimensional ($d=3$) systems in
this paper, although the application to the cases with $d\ne 3$ is straightforward.

\subsection{Landau-Gintzburg-Wilson action}
The Landau-Gintzburg-Wilson action for nonmagnetic fluctuations 
generally has odd order terms with respect to field variables, which 
is given as  
\begin{eqnarray}
S&=&
\sqrt{\frac{N}{T}}h\phi_{0}+\frac{1}{2}\sum_p a_p
 \phi_p\phi_{-p} \nonumber\\
&&+
 b\sqrt{\frac{T}{N}}\sum_{p_1p_2}\phi_{p_1}\phi_{p_2}\phi_{-p_1-p_2}\nonumber\\
&&+c\frac{T}{N}\sum_{p_1p_2p_3}\phi_{p_1}\phi_{p_2}\phi_{p_3}\phi_{-p_1-p_2-p_3}.\label{Action}
\end{eqnarray}
Here, $\phi_p$ is the nonmagnetic
field with $\phi_p^*=\phi_{-p}$, and $p$ represents the momentum and
energy. $N$ and $T$ are the number of
 lattice sites and temperature, respectively. 
In terms of $S$, the free energy $F$ is
given by $\exp(-F/T)=\int \mathcal{D}\phi \exp (-S)$. We neglect 
the momentum and energy 
dependence of the coefficients $b$ and $c$, which are irrelevant
from simple power counting in the context of the renormalization
group. For the stability of the system, $c>0$. 
$h$ is the field conjugate to the uniform
nonmagnetic field $\phi_0$; thus, does not depend on $p$.
Since we are interested in the orbital transition without breaking
translational symmetry, we use the Ornstein-Zernike form of $a_p$ as in the
ferromagnetic fluctuation\cite{Moriya1,Moriya2,MoriyaBook}:
\begin{eqnarray}
a_p=a_0+A|{\bf q}|^2+C|\omega_n|/|{\bf q}|, \label{ap}
\end{eqnarray}
where ${\bf q}$ and $\omega_n$ are the momentum and the Bosonic Matsubara
frequency, respectively. Existence of the odd order terms in eq. 
(\ref{Action}) leads to the first-order transition, CEP, and also
crossover, as shown in Fig. \ref{fig-phaseDMFT}.

\subsection{Modified SCR theory}
In the SCR theory, the non-Gaussian terms in the free energy $F$ are  
approximated by mean field decoupling.\cite{Moriya1,Moriya2,MoriyaBook,Misawa} This approximation corresponds 
to variationally determining the coefficients of the Gaussian
fluctuations.  
In our case with the presence of $h$ and $b$ terms, we need to
take into account the
order parameter itself in addition to mode-coupling terms. 
We will show that a modified Gaussian action can describe the
first-order transition, the CEP, and the crossover.

To describe the order parameter variations within the Gaussian
approximation, we assume the following variational action:
\begin{eqnarray}
S_{\rm eff}&=&\frac{1}{2}\sum_{p\ne 0}
 r_p\phi_p\phi_{-p}+\sqrt{\frac{N}{T}}\tilde{h}\phi_0+\frac{1}{2}r_0\phi_0^2,\\
r_p&=&a_p+\delta, 
\end{eqnarray}
where $\delta$ and $\tilde{h}$ are variational parameters. On the basis
of Gibbs-Bogoliubov-Feynman inequality, they are
determined by minimizing $\bar{\Omega}$\cite{textbook}:
\begin{eqnarray}
\bar{\Omega}&\equiv&{\Omega}_{\rm eff}+T\langle S-S_{\rm eff}\rangle,\\
\Omega_{\rm eff}&\equiv&-T\log Z_{\rm eff},\\
Z_{\rm eff}&\equiv&\int {\mathcal D}\phi \exp(-S_{\rm eff}),\\
&=&\Big(\prod_{p}\frac{1}{\sqrt{r_p}}\Big)
 \exp\Big(\frac{N}{T}\frac{\tilde{h}^2}{2r_0}\Big),\\
\langle S \rangle&\equiv&\int {\mathcal D}\phi S\exp(-S_{\rm
 eff})/Z_{\rm eff}.
\end{eqnarray}
Defining the order parameter $\phi$ as $\phi=\sqrt{T/N}\phi_0$, we obtain
\begin{eqnarray}
\frac{\bar{\Omega}}{N}&=&\frac{{\Omega}_{\rm eff}}{N}+(h-\tilde{h})\langle
 \phi\rangle-\frac{\delta}{2}\langle\phi^2\rangle+b\langle
 \phi^3\rangle+c\langle \phi^4\rangle\nonumber\\
&+&3(b\langle \phi\rangle+2c\langle \phi^2\rangle)\frac{T}{N}\sum_{p\ne
 0}\frac{1}{r_p}-\frac{\delta}{2}\frac{T}{N}\sum_{p\ne
 0}\frac{1}{r_p}\nonumber\\
&+&3c\Big(\frac{T}{N}\sum_{p\ne 0}\frac{1}{r_p}\Big)^2.
\end{eqnarray}
Since the effective action $S_{\rm eff}$ is Gaussian, we can
analytically calculate
$\langle \phi^n \rangle$ with $n$ being any integers, and they are given as
\begin{eqnarray}
\langle \phi\rangle&=&-\frac{\tilde{h}}{r_0},\label{phi1}\\
\langle \phi^2\rangle&=&\frac{T}{N}\frac{1}{r_0}+\frac{{\tilde{h}}^2}{r_0^2}=
\frac{T}{N}\frac{1}{r_0}+\langle \phi\rangle^2,\label{phi2}\\
\langle\phi^3\rangle&=&-\frac{T}{N}\frac{3\tilde{h}}{r_0^2}-\frac{{\tilde{h}}^3}{r_0^3}=
3 \langle \phi^2\rangle\langle \phi\rangle-2\langle \phi\rangle^3,\\
\langle\phi^4\rangle&=&-\frac{T^2}{N^2}\frac{3}{r_0^2}+\frac{T}{N}\frac{6{\tilde{h}}^2}{r_0^3}+\frac{{\tilde{h}}^4}{r_0^4}
=
3 \langle \phi^2\rangle^2-2\langle \phi\rangle^4.
\end{eqnarray}
Differentiating $\bar{\Omega}_{\rm eff}$ with respect to $\delta$, we obtain
\begin{eqnarray}
\lefteqn{\Big[ h-\tilde{h}+3b\frac{T}{N}\sum_{p\ne 0}\frac{1}{r_p}+3b\langle \phi^2 \rangle
-8c\langle \phi \rangle^3-6b\langle \phi \rangle^2
\Big]\frac{\partial \langle \phi \rangle}{\partial \delta}}\nonumber\\
&&+\Big[ -\frac{\delta}{2}+6c\frac{T}{N}\sum_{p\ne 0}\frac{1}{r_p}+3b\langle \phi \rangle
+6c\langle \phi^2 \rangle
\Big]\ \ \ \ \ \ \ \ \ \ \ \ \nonumber\\
&&\times \frac{\partial}{\partial\delta}\Big[\langle \phi^2
\rangle+\frac{T}{N}\sum_{p\ne 0}\frac{1}{r_p}\Big]=0.
\end{eqnarray}
Using eq. (\ref{phi2}) and defining 
\begin{eqnarray}
X\equiv\frac{T}{N}\sum_p\frac{1}{r_p}, \label{defX}
\end{eqnarray}
 we obtain 
\begin{eqnarray}
&&\Big[ h-\tilde{h}+3bX
-3b\langle \phi \rangle^2-8c\langle \phi \rangle^3
\Big]\frac{\partial \langle \phi \rangle}{\partial \delta}\ \ \ \ \ \ \
\ \ \ \ \ \ \ \ \ \ \nonumber\\
&&+\Big[ -\frac{\delta}{2}+6cX+3b\langle \phi \rangle
+6c\langle \phi \rangle^2
\Big] 
\frac{\partial}{\partial\delta}\Big[\langle \phi \rangle^2+X\Big]=0.\nonumber\\
\label{deldel}
\end{eqnarray}
Note that when $h=\tilde{h}=b=\langle \phi \rangle=0$,
eq. (\ref{deldel}) leads to the
self-consistent equation 
\begin{eqnarray}
\delta=12cX,
\end{eqnarray}
which coincides with the conventional expression.\cite{Moriya1,MoriyaBook} 
Similarly, differentiating $\bar{\Omega}_{\rm eff}$ with respect to
$\tilde{h}$ leads to 
\begin{eqnarray}
&&\Big[ h-\tilde{h}+3bX
-3b\langle \phi \rangle^2-8c\langle \phi \rangle^3
\Big]\frac{\partial \langle \phi \rangle}{\partial \tilde{h}}\ \ \ \ \ \ \nonumber\\
&&+\Big[ -\frac{\delta}{2}+6cX+3b\langle \phi \rangle
+6c\langle \phi \rangle^2
\Big] \frac{\partial}{\partial\tilde{h}}\langle \phi^2 \rangle=0.\label{delh}\ \ \ \ 
\end{eqnarray}
Combining eqs. (\ref{deldel}) and (\ref{delh}), we obtain
\begin{eqnarray}
\tilde{h}&=&h+3bX
-3b\langle \phi \rangle^2-8c\langle \phi \rangle^3,\label{sc-h}\\
\delta&=&12cX+6b\langle \phi \rangle
+12c\langle \phi \rangle^2.\label{sc-del}
\end{eqnarray}
Equations (\ref{sc-h}) and (\ref{sc-del}) are the self-consistent
equations in our theory.

Now, we derive  the equation
of state for $\langle \phi \rangle$. 
Using eqs. (\ref{phi1}), (\ref{sc-h}), and (\ref{sc-del}),  the equation
of state is derived as 
\begin{eqnarray}
h+3bX+(a_0+12cX)\langle \phi \rangle+3b\langle \phi \rangle^2+4c\langle \phi \rangle^3=0.\label{eq-state0}
\end{eqnarray}
Parameterizing $\langle \phi \rangle$ as $\langle \phi \rangle=\delta\phi-b/(4c)$
for later purposes, eq. (\ref{eq-state0}) is rewritten as 
\begin{eqnarray}
\Big[h-\frac{a_0b}{4c}+\frac{2b^3}{(4c)^2}\Big]+\Big(a_0-\frac{3b^2}{4c}+12cX\Big)\delta\phi
+4c\delta\phi^3=0.\label{eq-state1}
\end{eqnarray}
Equation (\ref{eq-state1}) is the basic equation used to determine the order
parameter $\delta\phi$. Note that $\delta\phi$ appears only in the  form of
$\delta\phi^2$ in $X$ from the definition of
$\delta\phi$ and eq. (\ref{sc-del}), and the effect of fluctuations on $h$ in
eq. (\ref{eq-state0}) is canceled out by the shift of $\phi\to\delta\phi-b/(4c)$. The
fluctuation term appears only
in the coefficient of $\delta\phi$ in eq. (\ref{eq-state1}). 
This equation describes the first-order transition, the CEP, and also the
crossover while taking into account the mode-coupling correction $X$
within the Gaussian fluctuation theory.
A similar 
equation of state was introduced in
ref. 28, where the renormalization of $h$ was neglected in eq. (\ref{eq-state0}).
 
\begin{figure}[t!]
	\begin{center}
    \includegraphics[width=0.5\textwidth]{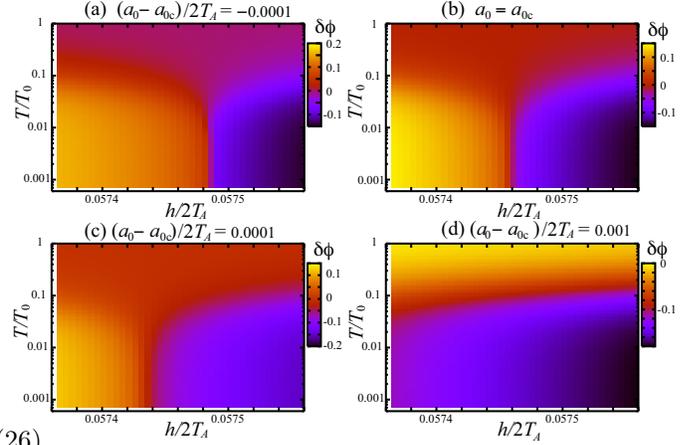}
\end{center}
\caption{(Color online) Contour plot of order parameter $\delta\phi$ in temperature-$h$
 plane for $b/2T_A=-0.01$, $c/2T_A=0.01$, $T_0=20$ K, $T_A=50$ K, and $x_c=300$. 
(a) $(a_0-a_{0c})/2T_A=-0.0001$, (b) $a_0=a_{0c}$, (c)
 $(a_0-a_{0c})/2T_A=0.0001$, and (d) $(a_0-a_{0c})/2T_A=0.001$. In (a), a finite temperature
 critical end point appears at $h/2T_A \sim 0.05749$ and $T/T_0\sim 0.01$, below
 which a first-order transition line exists. In (b), a
 zero-temperature critical end point exists at $h/2T_A\sim 0.05746$, where the
 order parameter $\delta\phi=0$ for $T\ge 0$. In (c) and (d), changes in the order
 parameter are smooth, indicating the crossover. }
\label{fig-phi}
\end{figure}

Figure \ref{fig-phi} shows the temperature and $h$ dependences of the order
parameter $\delta\phi$. Here, the critical value $a_{0c}$ is 
 given by eq. (\ref{a0c}) and the parameters $T_0$, $T_A$, and $x_c$
 introduced in Appendix
 \ref{detailcal} are fixed as $T_0=20$ K, $T_A=50$ K and $x_c=300$. 
In Fig. \ref{fig-phi}(a), there is a first-order
transition for $h/2T_A\sim 0.05749$ below $T/T_0\sim 0.01$. In
Fig. \ref{fig-phi}(b), the parameter $a_0$ is fixed as $a_0=a_{0c}$;
and thus, at zero temperature, the CEP is realized at $h\sim 0.05476$.
In Figs. \ref{fig-phi}(c) and \ref{fig-phi}(d), there is no phase transition and
$\delta\phi$ varies smoothly as functions of $T$ and $h$.

\subsection{Zero-temperature critical end point}\label{sec-zCEP}
In this subsection, we investigate the nonmagnetic orbital fluctuations near
the zero-temperature CEP in detail. Details of the calculations are summarized in Appendix
\ref{detailcal}.

Near the zero-temperature CEP, $r_0\sim 0$, 
$X$ in eq. (\ref{defX}) is expanded in terms of $r_0$ as 
\begin{eqnarray}
X=K_0+K_1r_0+\mathcal{K}(T), \label{exX}
\end{eqnarray}
with
\begin{eqnarray}
r_0&=&a_0+12c[K_0+K_1r_0+\mathcal{K}(T)]-\frac{3b^2}{4c}+12c\delta\phi^2,\nonumber\\
&=&\frac{1}{1-12cK_1}\Big[
a_0-\frac{3b^2}{4c}+12c\delta\phi^2+12c(K_0+\mathcal{K}(T))
\Big],\nonumber\\
\label{eq-r0}
\end{eqnarray}
and
\begin{eqnarray}
K_0>0,\ K_1<0, {\rm \ and \ } \mathcal{K}(T)>0.
\end{eqnarray}
As is well known, $\mathcal{K}(T) \propto T^{ 4/3 } $ in
three-dimensional systems with the dynamical exponent
$z=3$\cite{Moriya1,MoriyaBook} (see
Appendix \ref{detailcal}). 
Now, substituting eqs. (\ref{exX}) and (\ref{eq-r0}) into eq. (\ref{eq-state1}), we obtain
\begin{eqnarray}
H_0+\Big[A_0+A_1\mathcal{K}(T)\Big]\delta\phi
+4C_0\delta\phi^3=0,\label{eq-state3}
\end{eqnarray}
where
\begin{eqnarray}
H_0&=&h-\frac{a_0b}{4c}+\frac{2b^3}{(4c)^2},\label{eqH0}\\
A_0&=&a_0-\frac{3b^2}{4c}+\frac{12cK_1}{1-12cK_1}\Big[
a_0-\frac{3b^2}{4c}+\frac{K_0}{K_1}
\Big],\label{eqA0}\\
A_1&=&\frac{12c}{1-12cK_1}>0,\\
4C_0&=&4c\Big[1+\frac{36cK_1}{1-12cK_1}\Big].
\end{eqnarray}
In the case that the $\delta\phi$ dependence of $K_0$, $K_1$, and $\mathcal{K}(T)$ is negligible, 
eq. (\ref{eq-state3}) can be derived by minimizing the effective free
energy $F_{\rm eff}$ given as
\begin{eqnarray}
F_{\rm eff}=H_0\delta\phi+\frac{1}{2}\Big[A_0+A_1\mathcal{K}(T)\Big]\delta\phi^2
+C_0\delta\phi^4,
\end{eqnarray}
 which is nothing but the Landau's free energy for ferromagnetism in the
 presence of the magnetic field $-H_0$.\cite{Metamag}

Since $\mathcal{K}(T)$ vanishes at the  zero-temperature CEP as $T^{4/3}$,
the zero-temperature CEP is realized when $H_0=A_0=0$, which is equivalent to
$r_0=0$ and $\tilde{h}=0$. As a result, the critical value of the order
parameter is $\delta\phi=0$, leading to $\langle \phi \rangle_c=-b/(4c)$.
In terms of the original parameters, we obtain
the conditions for the zero-temperature CEP as
\begin{eqnarray}
a_{0c}&=&\frac{3b^2}{4c}-12cK_0,\label{a0c}\\
h_c&=&\frac{b^3}{(4c)^2}-3bK_0,\label{hc}
\end{eqnarray}
at zero temperature. 
Using eqs. (\ref{a0c}) and (\ref{hc}), eqs. (\ref{eqH0}) and
(\ref{eqA0}) are rewritten as 
\begin{eqnarray}
H_0&=&h-h_c,\\
A_0&=&(a_{0}-a_{0c})\frac{1+12cK_1}{1-12cK_1}.
\end{eqnarray}

Let us discuss the temperature dependence of $\delta\phi$ near the
zero-temperature CEP in the case of $C_0>0$ as expected
for a small $c>0$. 

First, we consider the case that all the parameters, $h$, $a_0$, $b$, $c$,
$A$, and $C$ are constant, {\it i.e.}, temperature-independent. Then, 
$H_0=0$ and $A_0=0$ in eq. (\ref{eq-state3}), 
when the CEP is realized at zero temperature. We obtain $\delta\phi=0$ or
\begin{eqnarray}
   \delta\phi^2=-\frac{A_1(0)}{4C_0(0)}\mathcal{K}(T), \label{x2}
\end{eqnarray}
where $A_1(0)$ and $C_0(0)$ are the zero-temperature values of $A_1$ and
$C_0$, respectively. Since $A_1(0)>0$, $\mathcal{K}(T)>0$, and
$C_0(0)>0$, the solutions of eq. (\ref{x2}) are pure imaginaries. 
 Thus, $\delta\phi=0$ for all temperatures 
  without taking into account the temperature
dependence in the parameters. This corresponds to the order parameter
variation in Fig. \ref{fig-phi}(b) for $h/2T_A\sim 0.05746$.  


Secondly, we consider that 
the leading temperature dependence of the parameters is proportional to 
$T^2$ as expected in Fermi liquid states. In this case, eq. (\ref{eq-state3})
 becomes 
\begin{eqnarray}
A_1(0)\mathcal{K}(T)\delta\phi+4C_0(0)\delta\phi^3\simeq-H^{''}_0T^2, \label{FLtemp}
\end{eqnarray}
where $H_0\simeq H^{''}_0T^2$. Thus, we obtain $\delta\phi\propto T^{2/3}$,
since this temperature dependence is self-consistent with
$\mathcal{K}(T)\propto T^{4/3}$.
Setting $\mathcal{K}(T)=\kappa T^{4/3}$ with $\kappa >0$, 
and $\delta\phi=\varphi T^{2/3}$, we obtain 
\begin{eqnarray}
A_1(0)\kappa \varphi+4C_0(0)\varphi^3\simeq -H^{''}_0.\label{coef-sc}
\end{eqnarray}
When $A_1(0)>0$ and $C_0(0)>0$, eq. (\ref{coef-sc}) always has
only one solution
$\varphi$, which is either positive or negative, depending on the sign of
$H_0^{''}$. The first-order transition occurs as $H''$ changes for
$A_1(0)<0$ and $C_0(0)>0$.

It is noted that $\phi$ couples any quantities and thus,
 in general, this nonanalytic temperature dependence $\delta\phi\propto
 T^{2/3}$ appears in any physical observables near the
 zero-temperature CEP. For example, the magnetic susceptibility
 $\chi_s$ has, rather than the usual $T^2$ dependence, the 
 nonanalytic temperature dependence as 
\begin{eqnarray}
\chi_s^{-1}(T)\simeq \chi_s^{-1}(0)+\lambda \delta\phi + \cdots.\label{chiS}
\end{eqnarray}
Here, $\lambda$ is the coupling constant between magnetic and
nonmagnetic fluctuations such as $\lambda \phi \sum_p M_pM_{-p}$ in the
action, where $M_p$ is the magnetic fluctuation at
 the momentum and the energy $p$. 
Thus, there might be a finite temperature region where 
$\chi_s(T)\propto T^{-2/3}$ is observed, when $\chi_s^{-1}(0)$ is
suppressed by some factors. Since the present theory is phenomenological,
the microscopic discussions on this issue are
 beyond the scope of this paper.

As for the other thermodynamic and transport quantities, 
we obtain the same critical behaviors as those in the ferromagnetic QCP with
$z=3$ and $d=3$. The specific heat coefficient diverges as $C/T\sim -\log
T$\cite{Makoshi}, and the resistivity is given as $\rho\sim T^{5/3}$,\cite{Ueda} in the asymptotic
limit. It is noted that, at a higher temperature, $\rho\propto T$ as is the
case for all the bosonic fluctuations. 

 
\subsection{Critical end point at finite temperature}
Here, we discuss the temperature dependence of $\delta\phi$ near a 
finite-temperature CEP at $T=T^*>0$.

As in the case of the zero-temperature CEP discussed in 
\S \ref{sec-zCEP}, we first consider 
the case that all the parameters have no temperature dependence.
In this case, $\delta\phi=0$ for $T>T^*$ and 
$\delta\phi\propto\sqrt{\mathcal{K}(T^*)-\mathcal{K}(T)}$ for $T<T^*$. Thus, we obtain 
$\delta\phi\propto (T^*-T)^{1/2}$ near $T\sim T^*$, 
exhibiting the mean field behaviors. This is understood
by noting that our theory is essentially mean field approximation and
that the $H_0=0$ line in the parameter space corresponds to the
case of the Ising model without magnetic fields, identifying 
$A_0+A_1\mathcal{K}(T)$ as $T-T_c$, where $T_c$ is the magnetic transition
temperature.

In the second case, the temperature dependence of all the parameters
is proportional to $T^2$ as derived in Fermi liquid states. Expanding the temperature dependence from
the critical end point temperature $T^*$, we obtain
\begin{eqnarray}
0&\simeq& \frac{\partial H_0}{\partial T}\Bigg|_{T=T^*}(T-T^*)+4C_0(T^*)\delta\phi^3.\label{first-finiteT}
\end{eqnarray}
Here, we assume $T^*\ll T_{\rm FL}$, where $T_{\rm FL}$ is the so-called Fermi
liquid temperature below which the Fermi liquid behaviors appear. 
In eq. (\ref{first-finiteT}), the temperature dependence in the second
term in eq. (\ref{eq-state1}) or (\ref{eq-state3}) is on the higher
order in $(T-T^*)/T^*$, and thus we neglect it. 
Then, the asymptotic temperature dependence of the order parameter $\delta\phi$ 
near the finite-temperature CEP is
\begin{eqnarray}
\delta\phi=\pm\Bigg| \frac{1}{4C_0(T^*)}\frac{\partial H_0}{\partial T}
\Bigg|_{T=T^*} \Bigg|^{\frac{1}{3}}|T^*-T|^{\frac{1}{3}}. 
\end{eqnarray}
Comparing this with the result at the zero-temperature CEP as discussed
in Sec. \ref{sec-zCEP}, one can see that 
there is a classical-quantum crossover from $\delta\phi\propto |T-T^*|^{1/3}$ to 
$\delta\phi \propto T^{2/3}$ as $T^*$ decreases in the second case.

Physically, it is unrealistic to consider $H_0=0$ for all temperatures in
real materials; thus, the second case is more realistic for
experimental situations. However, the first case is pedagogically
important, since it clearly exhibits this transition belonging to the Ising-type
universality class as expected. It is also important to note that 
$\delta\phi\propto T^{2/3}$ is realized only when $T_{FL}$ is well
defined, {\it i.e.,} when the Fermi liquid theory is valid and the
condition $T/T_{FL}\ll 1$ is realized.

\section{Discussion and Summary} \label{sec-sum}
We have discussed orbital fluctuations in a two-orbital Anderson lattice
model on the basis of the DMFT in \S \ref{sec-dmft} 
and constructed a self-consistent critical
theory of the orbital fluctuations in \S \ref{sec-scr}. 

In the first part of this paper in \S \ref{sec-dmft}, 
we have investigated meta-orbital
transition in the two-orbital Anderson model with anisotropic intersite
orbital-orbital interactions, and the schematic phase diagram is shown in
Fig. \ref{fig-phaseDMFT} as the summary of \S {\ref{sec-dmft}}. 

First, our results demonstrate that a  
 meta-orbital transition or crossover with steep changes in orbital
 occupancy
 occurs as pressure increases, only when the
 hybridization between the conduction electrons and the 
  f-electrons at ambient pressure is sufficiently small. 
For simplicity, let us consider that the pressure affects the 
 hybridizations only. Evidently, if
 the ambient pressure is located on the right side of the first-order
 transition line in Fig. {\ref{fig-phaseDMFT}}, the meta-orbital transition and
 crossover do not occur as pressure increases, since $v_B$ 
 increases as pressure increases. On the other hand, if the ambient
 pressure is located on the left side of the first-order transition line,
 it is possible to realize them, and if the parameters are tuned, the CEP 
 is realized. Here, we note that the f-electron 
ground state multiplet on the right
 side of the first-order transition line is the $A$-orbital, while it is
 the $B$-orbital on the left side of the line. 
Thus, it is concluded that the meta-orbital transition or
 crossover is realized only when the hybridization between the
 conduction electrons and the ground state multiplet of f-electrons at
 ambient pressure is {\it smaller } than that between the conduction
 electrons and the excited CEF multiplet.
These observations are valid when the dominant effect of the
 pressure is the variation in $v_{A,B}$. 

As far as we know, there is no compound that shows
a first-order  meta-orbital transition as pressure increases. This might
indicate that the ferro-orbital interaction is small in the real
materials or transverse orbital, and other interactions smear out the
first-order transition. 
As mentioned in \S \ref{sec-phase}, it is
important to clarify whether the first-order transition occurs in the
model without $J_o^z$. Since $J_o^z$ has been phenomenologically introduced to
incorporate the orbital-orbital interactions in the DMFT calculations, 
we need to
investigate the $J_o^z=0$ model on the basis of more elaborate methods
that can take into account both the local and intersite interactions.

Another important result shown in \S \ref{sec-dmft} is that 
the meta-orbital transition or crossover affects 
the f-electron occupancy. As shown in Fig. \ref{fig-Nf-Nc}, f-electron
occupancy slightly changes, when the meta-orbital transition or crossover 
occurs. This is understood by noting the hybridizations $v_A \ne
v_B$. This contrasts with the large valence change discussed in the
context of the valence transition driven by the large Coulomb
repulsion $U_{\rm fc}$, where the f-electron occupation number $n_{\rm f}$ 
cannot be close to 1, when the CEP is realized. The large $U_{\rm fc}$ also affects the
conduction electrons. Our recent study of the Ex-PAM based on DMFT+NRG calculations
reveals that, when the valence crossover occurs 
as pressure increases for the large $U_{\rm fc}$ comparable to the bandwidth,
the resulting density of states of the conduction electrons near the
crossover point shows 
a Hubbard-like peak deep below the Fermi level and a sharp peak near
the Fermi level,  which indicates 
 the existence of strongly correlated conduction electrons.

In the second part of this paper in \S \ref{sec-scr}, we have
developed a self-consistent renormalization theory of the orbital 
fluctuations. There, we have paid special attention to the order
parameter dependence of the mode-coupling term and derived self-consistent
equations, which lead to an equation of state for the order parameter. 
We have succeeded in describing the first-order transition and crossover
including the CEP of the orbital fluctuations. The temperature
dependence of the order parameter $\propto T^{2/3}$ is consistent with
that in the Kondo-volume-collapse theory\cite{KVC} and possibly consistent with that in
the valence transition\cite{MiyakePrivate}, since the symmetries of these order
parameters are the same. 

As for the temperature dependence of the resistivity, it is shown\cite{Holmes} that the valence
fluctuations lead to the $T$-linear dependence near the CEP, assuming a small $A$ in
eq. (\ref{ap}) from the results of the slave boson theory\cite{Onishi}. 
It is noted that the $T$-linear temperature dependence is realized in all the bosonic
fluctuations at high temperatures. For example, phonon
scatterings lead to such a behavior at high temperatures. In the case of
the orbital fluctuations, it is not clear whether the 
$A$ coefficient in
eq. (\ref{ap}) is sufficiently small in the present analysis. The 
 precise determination of the $A$ coefficient needs a more quantitative analysis. 
When $A\sim 0$ in eq. (\ref{ap}), 
we obtain $\mathcal{K}(T)\propto T^{2/3}$ near 
the CEP\cite{MiyakePrivate} and the equation of state (\ref{eq-state3})
leads to $\delta\phi\propto
T^{4/3}$ when all the parameters vary as a function of $T^2$ at low temperatures. 
Note that this temperature dependence 
should be cut off at the scale $T_A$, below which the asymptotic
behaviors characteristic to $d=3$ and $z=3$ discussed above appear.

In our theory, it is expected that 
CeCu$_2$Si$_2$
and related compounds and CeAl$_2$ at ambient pressure 
are on the left side of the line of the 
 first-order transition in Fig. \ref{fig-phaseDMFT}, {\it i.e.}, in the 
 $B$-rich phase. 
As for superconductivity, it is expected that compounds that show 
superconductivity away from their magnetic QCP pass near the CEP in
 Fig. \ref{fig-phaseDMFT} as pressure increases, while
 nonsuperconducting materials, such as CeAl$_2$, are not sufficiently close to
 the CEP, assuming that magnetic phases exist on the left side in the phase
 diagram in Fig. \ref{fig-phaseDMFT}. Since there are many aspects that affect the
 realization of superconductivity,  
such as the energy scale of the fluctuations and band
 structures\cite{Ikeda}, more quantitative model and analysis are needed in order to 
clarify the details of the superconductivity realized in these systems.

In summary, we have investigated orbital fluctuations in heavy-fermion
systems. We have shown that a meta-orbital transition occurs as pressure
increases and the orbital variations affect the density of states at the
Fermi
level and also f-electron occupancy. We have also constructed an effective
critical theory for the orbital fluctuations, which describes both the first-order transition and crossover
on the basis of a variational principle. Our results give a possible scenario for the
superconductivity and non-Fermi liquid behaviors associated with orbital
fluctuations in heavy-fermion systems.

\section*{Acknowledgment}
The author would thank H. Tsunetsugu, K. Miyake, and Y. Nishida 
for useful discussions. This work was supported by a Grant-in-Aid for
Scientific Research (No. 20740189) from the Japan Society for the
Promotion of Science. 

\appendix

\section{Localized limit and intersite interactions}\label{Kondolimit}
In this appendix, we will derive the intersite orbital interaction via a 
fourth-order perturbation in $v_{\alpha}$ from the localized limit of the
Hamiltonian (\ref{H}) without the $J_o^z$ term. Note that $J_0^z$ is not
exactly the same as the orbital intersite interaction that will be
derived in this appendix. $J_o^z$ has been introduced phenomenologically
in order to take into account the intersite orbital correlation in the DMFT
calculation in \S \ref{sec-dmft}. 

Let us consider Hamiltonian (\ref{H}) without $J_o^z$. 
In the limit of $U_{\alpha},\ U'\to \infty$, and 
$\varepsilon_{f\alpha}\to -\infty$
keeping $U_{\alpha}/|\varepsilon_{f\beta}| \gg 1$ and $v_\alpha
v_{\beta}/\varepsilon_{f\gamma}$ finite, the local excited 
configuration of f$^0$
 contributes to the low-energy properties of this system 
only as virtual states and configurations with f$^n(n\ge 2)$ are 
 neglected. Thus, 
 the effective Hamiltonian can be written by the localized spin and
 orbital degrees of freedom interacting with conduction electrons via
 exchange interactions: 
\begin{eqnarray}
H_{\rm ex}&=&\sum_{ij\sigma}(-t_{ij}-\mu\delta_{ij})c^{\dagger}_{i\sigma}c_{j\sigma}+\sum_{i \alpha \sigma\sigma'} J_{\alpha}
c^{\dagger}_{i\sigma}({\bf s}^c)_{\sigma\sigma'}
c_{i\sigma'}\cdot{\bf S}_{i}\nonumber\\
&+&\Delta\sum_i T_{iz}+\sum_{i \sigma\sigma'}
 J_{AB}c^{\dagger}_{i\sigma}({\bf s}^c)_{\sigma\sigma'}c_{i\sigma'}
T_{ix}\cdot{\bf S}_i\nonumber\\
&+&K^z_{AB}\sum_i n_{ic}T_{iz}
+K^{\pm}_{AB}\sum_i n_{ic}T_{ix},
\end{eqnarray}
where ${\bf S}_i$ represents the f-electron spin operator 
 at site $i$ and the conduction electron spin is given as 
$s^c_{\mu}=\sigma_{\mu}/2$, where
$\sigma_{\mu}$ with $\mu=x,y$ or $z$ is the Pauli matrix. $n_{ic}$ is
the conduction electron charge operator $n_{ic}=\sum_{\sigma}c^{\dagger}_{i\sigma}c_{i\sigma}$.
As for the orbital degrees of freedom of f-electrons, 
$T_{ix}=\sigma_x$ and $T_{iz}=\sigma_z$ acting on the
orbital space at site $i$ and the two eigenvalues of $T_{iz}$ 
are $1$ for the $A$-orbital 
and $-1$ for the $B$-orbital. 
The coupling constants are given as 
\begin{eqnarray}
J_{\alpha}&=&\frac{v_{\alpha}^2}{-\varepsilon_{f}},\\
J_{AB}&=&4K^{\pm}_{AB}=\frac{2v_Av_B}{-\varepsilon_{f}},\\
K^z_{AB}&=&\frac{v_A^2-v_B^2}{-4\varepsilon_{f}}.
\end{eqnarray}
Here, we assume $\varepsilon_{fA}\sim\varepsilon_{fB}\equiv \varepsilon_f$ and ignore
the difference between $\varepsilon_{fA}$ and $\varepsilon_{fB}$ in the denominator
of $J_{\alpha}$, $J_{AB}$, $K_{AB}^z$, and $K_{AB}^{\pm}$. 
The difference is taken into account in $\Delta=\varepsilon_{fA}-\varepsilon_{fB}$. 

Now, intersite interactions in the second-order perturbations of
$J_{\alpha}$, $J_{AB}$, $K_{AB}^z$, and $K_{AB}^+$ are obtained 
straightforwardly. Among them, the Ising-type orbital-orbital interaction
is given as
\begin{eqnarray}
{K_{AB}^z}^2\sum_{i,j}T_{iz}\chi_{ij}^cT_{jz},
\end{eqnarray}
where $\chi_{ij}^c$ is the conduction electron charge susceptibility.
Note that, when $v_B\gg v_A$ (or $v_A\gg v_B$), the coupling constant $J_B$
(or $J_A$) and $|K_{AB}^z|$ is much larger than $J_{AB}$ and $K_{AB}^+$.
In this limit, the spin-spin and Ising orbital interactions on the
fourth order in $v_{\alpha}$ are more important than those in the other sectors, such as the transverse
orbital and spin-orbital coupled interactions.

\section{Detailed calculations in SCR theory}\label{detailcal}
In this appendix, we summarize the detailed calculations of the mode
coupling correction $X$ in the SCR theory
in order to make this paper self-contained.  

Let us parameterize $X$ in eq. (\ref{defX}) as
\begin{eqnarray}
X=\frac{T}{N}\sum_{{\bf q},n}\frac{1}{r_0+A|{\bf q}|^2+C|\omega_n||{\bf q}|^{-\theta}},\label{eq-K}
\end{eqnarray}
with $\theta=1$.
Following the discussions by Misawa {\it et al.}\cite{Misawa}, eq. (\ref{eq-K}) is reduced
to
\begin{eqnarray}
X\!\!\!\!\!\!&=&\!\!\!\!\!\!\mathcal{K}(0)+\mathcal{K}(T),\\
\mathcal{K}(0)\!\!\!\!\!\!&=&\!\!\!\!\!\!\frac{K_d}{\pi}\int_0^{q_c}\!\!\!dq \int_0^{\infty}\!\!\!d\omega
\frac{C\omega q^{d+\theta-1}}{[q^{\theta}(r_0+Aq^2)]^2+(C\omega)^2},\label{Kappa0}\\
\mathcal{K}(T)\!\!\!\!\!\!&=&\!\!\!\!\!\!\frac{2K_d}{\pi}\int_0^{q_c}\!\!\!dq \int_0^{\infty}\!\!\!d\omega
\frac{n_B(\omega)C\omega
q^{d+\theta-1}}{[q^{\theta}(r_0+Aq^2)]^2+(C\omega)^2},\ \ \ \ \ \ \ \ \ \label{KappaT}
\end{eqnarray}
where $n_B$ is the Bose distribution function $n_B(\omega)$ $=$ $1/(e^{\omega/T}-1)$ and
$K_d$$=$$2\pi^{d/2}/[(2\pi)^d\Gamma(d/2)]$, with $\Gamma$ being the gamma function.
In terms of dimensionless parameters, eqs. (\ref{Kappa0}) and
(\ref{KappaT}) are rewritten as
\begin{eqnarray}
\mathcal{K}(0)\!\!\!\!\!\!&=&\!\!\!\!\!\!\frac{T_{0} d}{T_{A}}\int_0^{\infty}\!\!\!\!dz'z'\int_0^{x_c}\!\!\!\!
\frac{x^{d+\theta-1}dx}{[x^{\theta}(y+x^2)]^2+{z'}^2},\label{Kappa0-2}\\
\mathcal{K}(T)\!\!\!\!\!\!&=&\!\!\!\!\!\!\frac{2T_{0}d}{T_{A}}\int_0^{\infty}\!\!\!\!\frac{zdz}{e^{2\pi
 z}-1} \int_0^{x_c}\!\!\!\!
\frac{x^{d+\theta-1}}{[x^{\theta}(y+x^2)/t]^2+z^2}.\ \ \ \ \ \ \label{KappaT-2}
\end{eqnarray}
Here, $y=r_0/2T_A=(a_0+12cX+6b\langle \phi\rangle+12c\langle\phi \rangle^2)/2T_A$, $t=T/T_0$,
$T_A=Aq_B^2/2$, and $T_0=Aq_B^{2+\theta}/2\pi C$, where $q_B$ is the
 wave vector at the Brillouin zone boundary.

Near the critical end point $y=r_0/2T_A\sim 0$, eq. (\ref{Kappa0-2}) can be
expanded in terms of $y$, and we obtain $\mathcal{K}(0)=K_{0}+K_{1}r_0$.
The $z$ integral in eq. (\ref{KappaT-2}) can be carried out
analytically, and 
we obtain
\begin{eqnarray}
\mathcal{K}(T)\!\!\!\!\!\!&=&\!\!\!\!\!\!\frac{T_{0}d}{T_{A}}\int_0^{x_c}\!\!\!\!\!
dxx^{d+\theta-1}\Big[{\rm log}\ u -\frac{1}{2u}-\psi(u)\Big],\label{KappaT-3}
\end{eqnarray}
where $\psi$ is the di-gamma function and $u=x^{\theta}(y+x^2)/t$.
Rescaling $x$ by $t^{1/(2+\theta)}$, eq. (\ref{KappaT-3}) becomes
\begin{eqnarray}
\mathcal{K}(T)\!\!\!\!\!\!&=&\!\!\!\!\!\!\frac{T_{0}d}{T_{A}}t^{\frac{d+\theta}{2+\theta}}
\int_0^{x_c/t^{\frac{1}{2+\theta}}}\!\!\!\!\!\!\!\!\!\!\!\!\!\!\!ds
s^{d+\theta-1}\Big[{\rm log}\ u -\frac{1}{2u}-\psi(u)\Big].\ \ \ \ \ \ \label{KappaT-4}
\end{eqnarray}
Here, $u=\alpha(y,t,\theta)s^{\theta}+s^{2+\theta}$ with
$\alpha(y,t,\theta)=y/t^{\frac{2}{2+\theta}}$. As discussed by Misawa 
{\it et al}.,\cite{Misawa} eq. (\ref{KappaT-4}) is proportional to
$t^{\frac{d+\theta}{2+\theta}}$ if $\alpha(y,t,\theta)\to 0$ for
$t\to 0$. This is understood by noting that $u\to s^{2+\theta}$ for
$t\to 0$ and the integral has no $t$ dependence for a small $t$. 

\end{document}